# Electron-induced chemistry in imidazole clusters embedded in helium nanodroplets

M. Kuhn, S. Raggl, P. Martini, N. Gitzl, M. Mahmoodi Darian[1], M. Goulart, J. Postler, L. Feketeová[2,a], P. Scheier[b]

Institut für Ionenphysik und Angewandte Physik, Universität Innsbruck, Technikerstraße 25, 6020 Innsbruck, Austria.

[1] Department of Physics, Karaj Branch, Islamic Azad University, Karaj, Iran.

[2] Université de Lyon; Université Claude Bernard Lyon1; CNRS/IN2P3, UMR5822, Institut de Physique Nucléaire de Lyon, 43 Bd du 11 novembre 1918, 69622 Villeurbanne, France.

Correspondence and PROOFS to:

[a] Dr. Linda Feketeová: E-mail: l.feketeova@ipnl.in2p3.fr

[b] Prof. Paul Scheier: E-mail: paul.scheier@uibk.ac.at

**Abstract**

Electron-induced chemistry in imidazole (IMI) clusters embedded in helium nanodroplets (with an average size of $2\times10^5$ He atoms) has been investigated with high-resoluton time-of-flight mass spectrometry. The formation of both, negative and positive, ions was monitored as a function of the cluster size $n$. In both ion spectra a clear series of peaks with IMI cluster sizes up to at least 25 are observed. While the anions are formed by collisions of $IMI_n$ with $He^{*-}$, the cations are formed through ionization of $IMI_n$ by $He^+$ as the measured onset for the cation formation is observed at 24.6 eV (ionization energy of He). The most abundant series of anions are dehydrogenated anions $IMI_{n-1}(IMI-H)^-$, while other anion series are IMI clusters involving CN and $C_2H_4$ moieties. The formation of cations is dominated by the protonated cluster ions $IMI_nH^+$, while the intensity of parent cluster cations $IMI_n^+$ is also observed preferentially for the small cluster size $n$. The observation of series of cluster cations $[IMI_nCH_3]^+$ suggests either $CH_3^+$ cation to be solvated by $n$ neutral IMI molecules, or the electron-induced chemistry has led to the formation of protonated methyl-imidazole solvated by $(n-1)$ neutral IMI molecules.

**Running Title:** Electron-induced chemistry in imidazole doped HND.

# 1 Introduction

Imidazole (IMI) $C_3H_4N_2$ is ubiquitous in nature as an important building block in biology (Scheme 1). It is present in, for example, the amino acid histidine, the hormone histamine, or antifungal drugs and antibiotics, and when fused with pyrimidine it forms purine that is the most widely distributed nitrogen containing heterocyclic molecule in nature [1]. Thus, IMI belongs to the simplest heterocyclic compounds of a five-membered aromatic ring containing two nitrogen atoms. Electron-induced processes in gas-phase IMI, such as, electron ionization and dissociative electron attachment (DEA), have been well documented [2,3]. While the ionization of IMI leads predominantly to the release of neutral HCN associated with the ring opening, and to the loss of an H atom [2], the DEA to IMI is dominated by H atom loss [3]. However, other DEA reactions were observed associated with the opening of the ring structure, including multiple dehydrogenation reactions [3]. Many of the anions formed through DEA to IMI are associated with the release of neutral HCN. Thus, the release of HCN seems to be an important by-product in the decomposition of both, positive and negative ions of IMI molecule [2,3].

**[Insert Scheme 1 here, please]**

The fragmentation induced by energetic particles may change significantly going from isolated molecules to molecular clusters [4], including an opening up of new fragmentation channels [4,5] and/or new covalent bond formation [6,7]. Recent work on ionization of neutral IMI clusters by electrons and photons has pointed out the strong hydrogen bonds N−H⋯N in IMI, which can lead to hydrogen or proton transfer between the cluster constituents upon excitation, resulting in the observation of protonated IMI clusters $(C_3H_4N_2)_nH^+$ with n ≤ 5 [8]. *Ab initio* calculations on neutral IMI clusters suggest the hydrogen bond N−H⋯N structural motif to be always present [8-10], which was confirmed in the case of the neutral dimer by rotational spectroscopy [10]. The recent infrared spectroscopy study of the IMI trimer in helium nanodroplets has shown through unambiguous free NH stretching mode that imidazole molecules tend to form linear structures due to their dipole-dipole interactions [11].

Superfluid helium nanodroplets (HND) offer a unique environment to study chemical reactions at ultracold temperatures [12-24]. Atoms and molecules that collide with a droplet are picked up and transferred in most cases to the centre of the droplet. Evaporation of weakly bound He atoms cools quickly both the droplet and the dopants down to 0.37 K [25]. The existence of metastable excited states in superfluid HND can trigger the ionization of embedded atoms and molecules by Penning ionization [26]. These metastable states can be promoted, e.g., by the inelastic collision of a He atom with an electron. Additionally, an inelastically scattered electron may recombine with the metastable He atom forming highly mobile $He*^-$ that can transfer its negative charge and/or electronic excitation energy to the dopant [16,27]. This process showed to be dominant at electron energies of ~ 22 eV [27]. On the other hand, the cations of dopants in HND can be produced also via formation of $He^+$ that approaches the dopant via resonant hole-hopping and a final highly exothermic charge transfer to the dopant [28]. The cooling power of superfluid He has been reported to quench many fragmentation reactions and thus enables the study of short-lived reaction intermediates [29]. Here we utilize a high-resolution time-of-flight mass spectrometry to study formation of positive and negative ions upon electron collisions with helium nanodroplets doped with IMI.

# 2 Experimental section

He nanodroplets are formed upon expansion of pressurized (2.4MPa) and precooled (9.75K for cations and 9.80K for anions) He with a purity of 99.9999% through a 5µm pinhole into ultra-high vacuum. Evaporative cooling drops the temperature of the droplets to 0.37K [25,30]. Under these conditions, the resulting droplet size distribution is log-normal with an average size of about $2\times10^5$ He atoms [31]. After passing a skimmer to prevent shock fronts that destroy the droplets they enter a differentially pumped pickup chamber where IMI vapor is present. IMI (purity $\geq$ 99%) was used as received from Sigma Aldrich (Austria). The sample was heated in a container outside the instrument to a temperature between 358K and 423K and introduced via a slightly higher heated (to prevent condensation) gas line and a valve into a heated pickup cell. The IMI doped HND pass into another differentially pumped vacuum chamber where they are crossed with an electron beam of variable energy. Cations are formed with typical electron energies above 70 eV whereas anions are formed at electron energies below 30 eV. The electron energy scale is calibrated via the onset of the ion efficiency curve of $He^+$ at 24.6eV [ionization energy (IE) of He] or the 22 eV resonance leading to the formation of $He^{*-}$ [32, 33]. Cluster ions that are ejected from the HND and most of the time free of attached He atoms are guided with electrostatic lenses to the extraction region of an orthogonal extraction reflectron time of flight mass spectrometer (Hi-TOF, TofWerk GmbH) with a mass resolution $m/\Delta m=3500$ in the present study. The particle density of the IMI was controlled by the heating power and set to achieve a desired cluster size distribution.

# 3 Results and discussion

## 3.1 Anion formation upon electron irradiation of imidazole doped HND

Figure 1 shows the mass spectrum for negative ions formed upon electron bombardment of HND doped with IMI at an electron energy of 22 eV.

**[Insert Figure 1 here, please]**

As already mentioned, the dominant process at this electron energy with He is the formation of highly mobile He*$^-$ (within superfluid HND) by the attachment of the scattered electron to the excited He* [27]. The ion-induced dipole interaction between the He*$^-$ and the dopant will attract the two species and lead to the transfer of the negative charge and/or electronic excitation energy. In dissociative electron attachment to gas-phase molecules each fragment anion is formed at specific resonances [34], however, in doped HND all possible anions are formed through the interaction of the dopant molecule with He*$^-$ [35]. In Figure 2 we show anion efficiency curves for some selected anions around the dimer and trimer region of the cluster size distribution. Two relatively weak features are observed at around 13 eV and 32 eV, however, the dominant resonance is at 22 eV corresponding to an intermediate He*$^-$ that is responsible for the formation of the anions observed in the present study at this electron energy.

In the Figure 1, several series of peaks spaced by the mass of the IMI molecule are visible in the spectrum with the dehydrogenated parent anions [IMI$_n$−H]$^-$, with *n* up to 25, being the most abundant anions in the spectrum. Other series of anions formed are highlighted in the inset of the Figure 1, which shows the enlarged region of the mass spectrum between dehydrogenated IMI trimer [IMI$_3$−H]$^-$ and dehydrogenated tetramer [IMI$_4$−H]$^-$. The product anions [IMI$_3$CN]$^-$ at *m/z* 230.1 and [IMI$_2$(IMI−H)C$_2$H$_4$]$^-$ at *m/z* 231.1 can be assigned, and three less abundant anions at *m/z* 256.1, 257.1 and 258.1 corresponding to [IMI$_n$(CN)$_2$]$^-$, [IMI$_{n-1}$(IMI-H)CNC$_2$H$_4$]$^-$ and [IMI$_n$CNH$_2$CN]$^-$, respectively.

**[Insert Figure 2 here, please]**

In order to deduce the correct yield of all anions contributing with their isotopic patterns to the measured mass spectrum we utilized our recently developed software IsotopeFit [36] and plotted in the following the evolution of the ion signal as a function of the cluster size *n*. Such a distribution for the formation of dehydrogenated anions [IMI$_{n+1}$−H]$^-$ as a function of the cluster size *n* is shown in Figure 3. The ion yield represents the total counts of the respective ion including all possible isotopologues determined by IsotopeFit [36]. The dehydrogenated anion [IMI$_{n+1}$−H]$^-$ can be also written in the form IMI$_n$(IMI−H)$^-$. It is likely that the structure contains the ionic core (IMI−H)$^-$ solvated by neutral IMI molecules. The formation of the dehydrogenated IMI anion (IMI−H)$^-$ is the most abundant product anion channel in dissociative electron attachment to isolates molecules of IMI in the gas phase [3]. The electron affinity of the (IMI−H)$^•$ radical is the highest (2.6 eV) when H is removed from the N1 position of IMI (see Scheme 1) [3,37]. For the formation of the anion IMI$_n$(IMI−H)$^-$, the precursor cluster had to be of the size ≥ *n+1*, where one of the IMI molecules fragmented upon the electron transfer from He*$^-$. The thermodynamic threshold for the formation of (IMI−H)$^-$ was reported to be 1.36 eV (measured in the DEA to gas-phase IMI [3]) slightly smaller than determined from another study 1.51 eV [37]. The endothermicity of the reaction suggest that some of the excitation energy has been transferred from He*$^-$ to the dopant as well.

**[Insert Figure 3 here, please]**

We also observe a formation of parent anions of IMI clusters as it was the case in a number of other biomolecules studied in HND, such as DNA bases [38] and amino acids [16,39]. The $IMI_n^-$ cluster series is shown in Figure 4a together with a section of the mass spectrum (Figure 4b) showing the fit of the experimental data. For some amino acids, the intensity of the parent anions can be comparable or even larger than the intensity of the corresponding dehydrogenated anion. However, in the case of IMI the intensity of $IMI_n^-$ cluster anions is three times smaller than the isotopologue of the dehydrogenated anions. The distribution shown in Figure 4a suggest higher cluster size $n$ can better stabilize formed radical anion $IMI^-$.

**[Insert Figure 4 here, please]**

Another abundant ions observed in the anion formation in IMI clusters embedded in HND is the series of anions $[IMI_nCN]^-$. The cluster size distribution of these anions is shown in Figure 5. The distribution shows notably high intensity for the formation of the anion $CN^-$, i.e., $n = 0$. The CN molecule has a very high electron affinity of 3.8 eV [40]. The possible formation of the anion $CN^-$ has been noted in the electron attachment studies to gas phase IMI [3], however, the intensity of this ion has been contaminated by the formation of $CN^-$ by DEA to another precursor molecule. Nevertheless, the high electron affinity of the CN molecule suggests that the cluster anions $[IMI_nCN]^-$ have an ionic core $CN^-$ solvated by neutral IMI molecules. In any case, the formation of $CN^-$ from IMI is a complex reaction that requires cleavage of the IMI ring by breaking of two bonds and at least another bond to remove the H atom from the liberating CN moiety. The thermodynamic threshold for the formation of CN⁻ in DEA to gas-phase IMI was calculated to be 2.01 eV [3], which is only slightly above the threshold for the formation of dehydrogenated anion discussed above.

**[Insert Figure 5 here, please]**

Another series of anions formed in IMI doped HNDs corresponds to IMI clusters with an extra mass of 23 amu. Thanks the high resolution of the mass spectrometer we are confident in assigning the extra mass to $C_2H_3$ moiety. Even though the EA of $C_2H_3$ is positive of 0.67 eV [40], the EA of (IMI−H) is much higher of 2.61 eV [37] and thus, we assign this cluster series to $[IMI_{n-1}(IMI-H)C_2H_4]^-$ anions. The decomposition of IMI molecule to form $C_2H_4$ + NCN can be estimated from the available data [40] to be endothermic by 4.1 eV, whereas the decomposition of two IMI molecules to form $C_2H_4$ + 4 HCN lowers the endothermicity to 3.5 eV. The corresponding cluster size distribution is shown in Figure 6. The distribution of $[IMI_{n-1}(IMI-H)C_2H_4]^-$ cluster anions has obviously a different shape than the above mentioned distribution for $[IMI_nCN]^-$ anions shown in Figure 5. The distribution of $[IMI_{n-1}(IMI-H)C_2H_4]^-$ has a bimodal character and at least three IMI molecules are required to form the smallest cluster anion. It is highly likely that the structure of these clusters has ionic core $(IMI-H)^-$ solvated by $(n-1)$ neutral IMI molecules and newly formed $C_2H_4$, thus, can be written as $IMI_{n-1}(IMI-H)^-C_2H_4$. Another interesting observation is the two pronounced intensity drops at $n=5$ and $n=10$, which are also noted but less pronounced for other anion cluster series discussed above.

**[Insert Figure 6 here, please]**

The formation of the smaller abundant anions mentioned at the beginning of this section that correspond to the series $[IMI_n(CN)_2]^-$, $[IMI_{n-1}(IMI-H)CNC_2H_4]^-$ and [IMI$_n$CNH$_2$CN]$^-$ (see the inset of the Figure 1) highlights much more complex reaction in IMI cluster upon electron uptake. As product ions containing C$_2$H$_4$ moieties exhibit a minimum cluster size we expect that at least two IMI monomers are involved in their formation process. As discussed above, the reaction between two IMI molecules lowers the endothermicity for the formation of C$_2$H$_4$ moiety associated with the formation of several HCN molecules. Attachment of the electron to HCN molecule can lead to the formation of CN$^-$ moiety noted in several of the anion cluster series observed.

### 3.2 Cation formation upon electron irradiation of imidazole doped HND

Figure 7 shows the mass spectrum of positively charged ions from HND doped with IMI molecules recorded at electron energy of 88 eV. All positively charged product ions from the IMI doped HND are formed at electron energies above 24.6 eV which indicates that charge transfer from an initially formed He$^+$ is the dominant process.

The most abundant series of peaks spaced by the mass of IMI molecule are protonated IMI cations IMI$_n$H$^+$ observed up to $n < 24$. Other cations observed in the spectrum correspond to parent cluster ions IMI$_n^+$ and cluster ions with an extra CH$_3$ group IMI$_n$CH$_3^+$. The inset of the Figure 7 highlights the formation of IMI$_3$CH$_3^+$ and the formation of IMI$_3$HCN$^+$.

**[Insert Figure 7 here, please]**

We have deduce the correct yield of all cations contributing with their isotopic patterns to the measured mass spectrum using the software IsotopeFit [36] and plotted in the following the evolution of the ion signal as a function of the cluster size $n$. Figure 8 shows the ion yield for the formation of the most abundant protonated imidazole cluster cations IMI$_n$H$^+$.

Protonated IMI clusters were the only ions observed in the electron- and photo-ionization of bare IMI clusters [8], however, with sizes only up to $n = 5$. The *ab initio* calculations in Poterya et al. [8] showed that ionization of an IMI dimer is followed by a proton transfer from one IMI to the other. Thus, in the formation of imidazole cluster cations $IMI_nH^+$ a radical $(IMI-H)^\bullet$ needs to be ejected from the HND.

**[Insert Figure 8 here, please]**

Another series of cations observed corresponds to the ionized imidazole clusters $IMI_n^+$, which distribution is shown in Figure 9. The distribution shows an exponential decay, however, the parent monomer ions $IMI^+$ may be formed also upon electron ionization of isolated IMI molecules reaching the ion source. The intensity of $IMI_n^+$ cluster cations decreases with increasing $n$, however, not as fast as the intensity of protonated cluster cations $IMI_nH^+$ shown in Figure 8. Thus, the larger the precursor cluster, the less likely it is to transfer a proton and to form the protonated cluster cation $IMI_nH^+$ than the parent cluster cation $IMI_n^+$. Nevertheless, the intensity of the parent cluster cation $IMI_n^+$ is always smaller than the respective protonated cluster $IMI_nH^+$.

**[Insert Figure 9 here, please]**

The most peculiar cations observed to be formed are cluster cations of the form $IMI_nCH_3^+$. The intensity of these cations as a function of the cluster size $n$ is shown in Figure 10. The formation of $CH_3^+$ ion through ionization of a sole IMI seems unlikely as the low mass fragment ion in the ionization of IMI [2] is appearing at *m/z* 14. Even though the origin of this mass is not discussed, the deuterated studies suggest this fragment to correspond to a $CH_2^+$ cation [2]. Therefore, it is more likely that the $CH_3^+$ cation is formed in the reaction of at least two IMI molecules. Considering the structure of the IMI$_n$CH$_3^+$ cluster cations, there are two possibilities, either the formed CH$_3^+$ cation is solvated by $n$ neutral IMI molecules, or the formed radical cation CH$_2^+$ reacts with another IMI molecule to form protonated methyl-imidazole C$_4$H$_7$N$_2^+$ that is solvated by ($n-1$) neutral IMI molecules.

[Insert Figure 10 here, please]

## 4 Conclusions

The formation of both, negative and positive, ions upon an electron collision in imidazole (IMI) clusters embedded in HND was investigated with high-resoluton time-of-flight mass spectrometry and was studied as a function of the cluster size $n$. In both ion spectra a clear series of peaks with IMI cluster sizes up to at least 25 are observed indicated by the ions spaced by the mass of the IMI molecule.

The anions are formed in the interaction of IMI$_n$ with He$^{*-}$ (formed in the collision of an electron with HND) as no resonances were observed up to 22 eV of electron energy. The most abundant series of anions are dehydrogenated anions IMI$_{n-1}$(IMI−H)$^-$, which likely have a structure of dehydrogenated anion (IMI−H)$^-$ being solvated by ($n-1$) neutral IMI molecules. Other anion series observed were IMI clusters involving CN and C$_2$H$_4$ moieties, such as, [IMI$_n$CN]$^-$, [IMI$_{n-1}$(IMI−H)C$_2$H$_4$]$^-$, or smaller abundant anions [IMI$_n$(CN)$_2$]$^-$, [IMI$_{n-1}$(IMI-H)CNC$_2$H$_4$]$^-$ and [IMI$_n$CNH$_2$CN]$^-$.

The measured onset for the cation formation is observed at 24.6 eV that is the IE of He, which indicates that the cations are formed through ionization of IMI$_n$ by He$^+$. The formation of cations is dominated by the formation of protonated cluster ions IMI$_n$H$^+$. The formation of parent cluster cations IMI$_n^+$ is also observed, however, preferentially for the small cluster size $n$, while the intensity of IMI$_n^+$ decreases with $n$ much more rapidly than in the case of the protonated IMI$_n$H$^+$ cluster cations. The observation of series of [IMI$_n$CH$_3$]$^+$ cluster cations suggests either CH$_3^+$ cation to be formed and solvated by $n$ neutral IMI molecules, or the formation of protonated methyl-imidazole solvated by ($n-1$) neutral IMI molecules.

**Acknowledgements:** This work was supported by the Austrian Science Fund FWF, Wien (P26635, and M1908). M.M.D. thanks the University of Innsbruck for support via P7440-016-028. L.F. is thankful for the support from the Vicerectorate for Research – University of Innsbruck via P7440-035-011, the Institut de Physique Nucleáire de Lyon, and the support from the Labex-LIO (Lyon Institute of Origins).

**Scheme 1:** Chemical structure of imidazole.

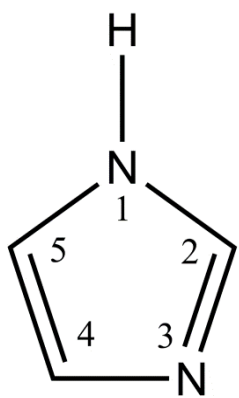

1-H-imidazole **1**

**Figure Captions:**

**Figure 1:** Mass spectrum of negatively charged ions from HND doped with imidazole: $p_{He}$=2.4MPa, $T_{He}$=9.8K, $E_{el}$=22eV, $I_{el}$=209µA, $T_{IMI}$=423K. The peaks designated by the triangles (red) and squares (green) correspond to $[IMI_n-H]^-$ and $[IMI_nCN]^-$ cluster series, respectively. The inset in the figure shows in detail the range between $[IMI_3-H]^-$ and $[IMI_4-H]^-$. The doublet at m/z=230.1 and 231.1 correspond to the anions $[IMI_3CN]^-$ and $[IMI_2(IMI-H)C_2H_4]^-$, respectively. In addition, the three weak peaks at m/z= 256.1, 257.1 and 258.1 can be assigned to $[IMI_3(CN)_2]^-$, $[IMI_2(IMI-H)CNC_2H_4]^-$ and $[IMI_3CNH_2CN]^-$, respectively.

**Figure 2:** Anion efficiency curves for selected anions formed via inelastic collisions of electrons with HNDs doped with IMI. Parameters: temperature and pressure of the He before expansion: 9.65K, 2.4MPa; $I_{el}$=30µA, $T_{IMI}$=423K.

**Figure 3:** Cluster size distribution of the dehydrogenated closed-shell anions, $[IMI_{n+1}-H]^-$ taken from the mass spectrum shown in Figure 1. $p_{He}$=2.4MPa, $T_{He}$=9.8K, $E_{el}$=22eV, $I_{el}$=209µA, $T_{IMI}$=423K. The ion yield represents the total counts of the respective ion, including all possible isotopologues and was determined via IsotopeFit [36].

**Figure 4:** (a) Cluster size distribution of the open shell parent anions, $IMI_n^-$ taken from the mass spectrum shown in Figure 1. $p_{He}$=2.4MPa, $T_{He}$=9.8K, $E_{el}$=22eV, $I_{el}$=209µA, $T_{IMI}$=423K. The ion yield represents the total counts of the respective ion, including all possible isotopologues and was determined via IsotopeFit [36]. (b) Section of a mass spectrum showing the ions $IMI_6^-$ (blue dashed line) and $[IMI_6-H]^-$ (red dashed line) fitted to the measured mass spectrum (grey line). The black line is the best fit of the two ions to the mass spectrum obtained via IsotopeFit [36].

**Figure 5:** Cluster size distribution of the ion series, [IMI$_n$CN]$^-$ taken from the mass spectrum shown in Figure 1. $p_{He}$=2.4MPa, $T_{He}$=9.8K, $E_{el}$=22eV, $I_{el}$=209μA, $T_{IMI}$=423K. The ion yield represents the total counts of the respective ion, including all possible isotopologues and was determined via IsotopeFit [36].

**Figure 6:** Cluster size distribution of the ion series, [IMI$_{n-1}$(IMI−H)C$_2$H$_4$]$^-$ taken from the mass spectrum shown in Figure 1. $p_{He}$=2.4MPa, $T_{He}$=9.8K, $E_{el}$=22eV, $I_{el}$=209μA, $T_{IMI}$=423K. The ion yield represents the total counts of the respective ion, including all possible isotopologues and was determined via IsotopeFit [36].

**Figure 7:** Mass spectrum of positively-charged ions from HND doped with imidazole: $p_{He}$=2.4MPa, $T_{He}$=9.75K, $E_{el}$=88eV, $I_{el}$=82μA, $T_{IMI}$=358K. The peaks designated by the triangles (red) and by the stars (blue) correspond to IMI$_n$H$^+$ and IMI$_n$CH$_3^+$ cluster series, respectively. The peak designated with an open circle in the inset originates from IMI$_3$HCN$^+$. The peak four mass units to the right can be assigned as IMI$_3$H$_3$O$^+$ and originates from the additional pickup of a water molecule present in the residual gas of the instrument. The peaks series separated by four mass units originates from He$_n^+$ ions.

**Figure 8:** Cluster size distribution of the protonated imidazole ions, IMI$_n$H$^+$, taken from the mass spectrum shown in Figure 7: $p_{He}$=2.4MPa, $T_{He}$=9.75K, $E_{el}$=88eV, $I_{el}$=82μA, $T_{IMI}$=358K. The ion yield represents the total counts of the respective ion, including all possible isotopologues and was determined via IsotopeFit [36].

**Figure 9:** Cluster size distribution of the parent imidazole ions, $IMI_n^+$, taken from the mass spectrum shown in Figure 7: $p_{He}$=2.4MPa, $T_{He}$=9.75K, $E_{el}$=88eV, $I_{el}$=82µA, $T_{IMI}$=358K. The ion yield represents the total counts of the respective ion, including all possible isotopologues and was determined via IsotopeFit [36]. Please note that the parent monomer ions $IMI^+$ may be formed additionally upon electron ionization of isolated imidazole molecules reaching the ion source.

**Figure 10:** Cluster size distribution of the ions, $IMI_nCH_3^+$, taken from the mass spectrum shown in Figure 7: $p_{He}$=2.4MPa, $T_{He}$=9.75K, $E_{el}$=88eV, $I_{el}$=82µA, $T_{IMI}$=358K. The ion yield represents the total counts of the respective ion, including all possible isotopologues and was determined via IsotopeFit [36].

**Figure 1**

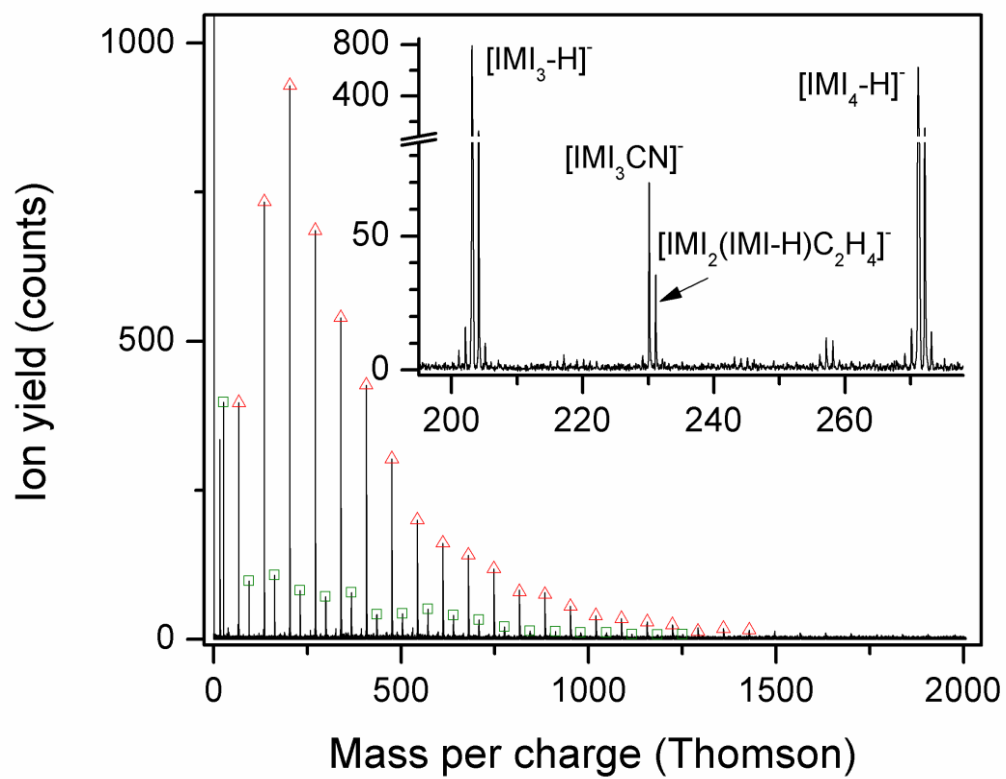

**Figure 2**

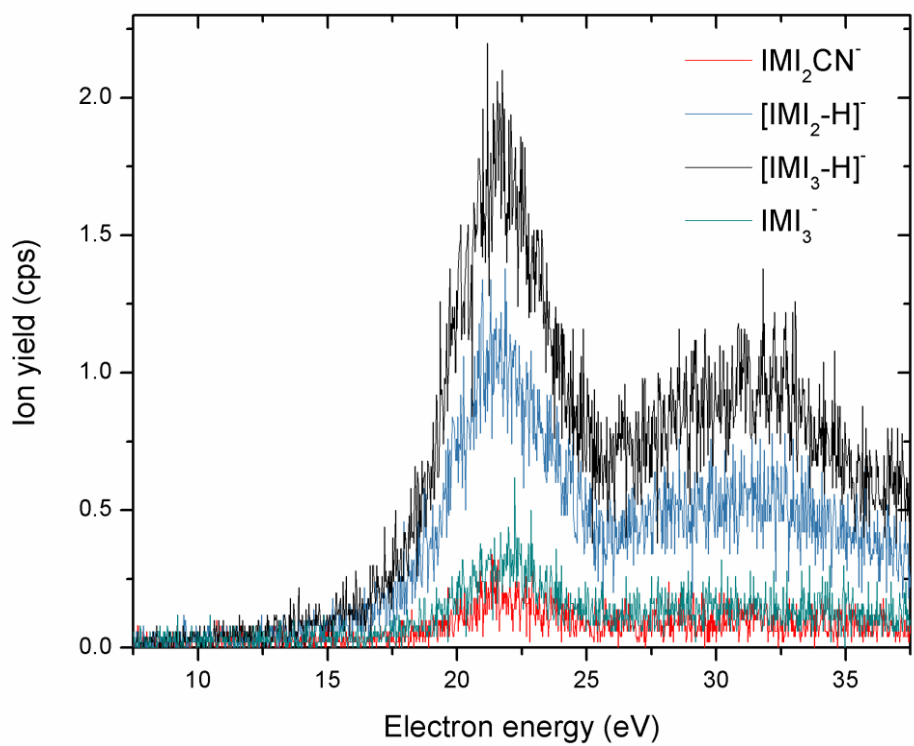

**Figure 3**

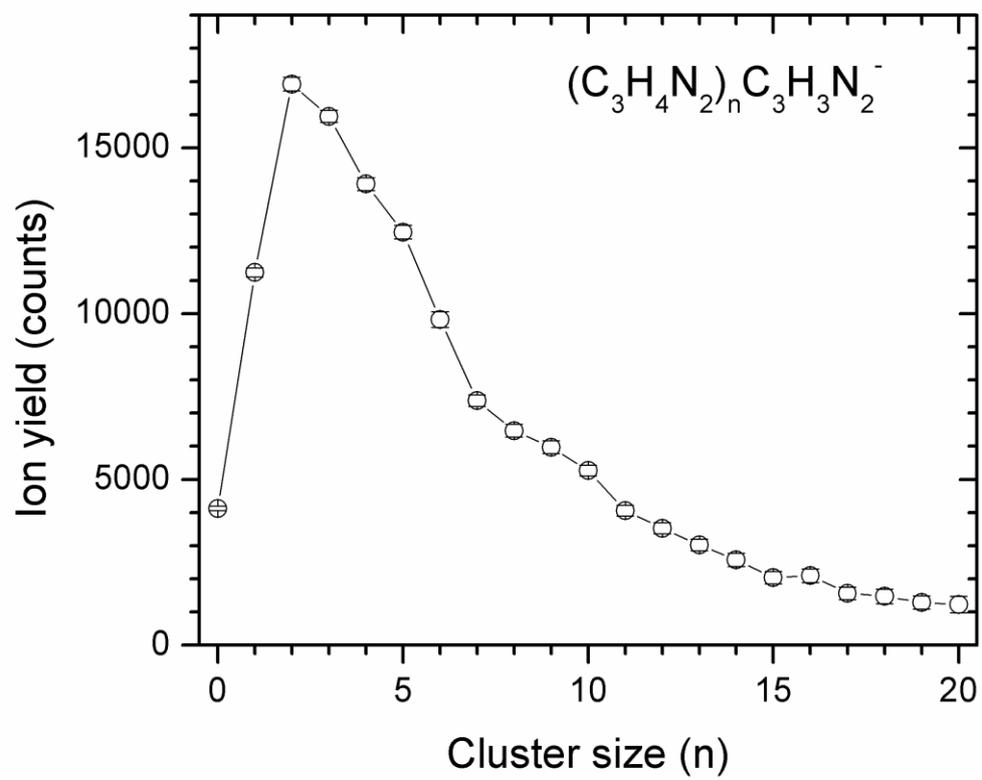

**Figure 4**

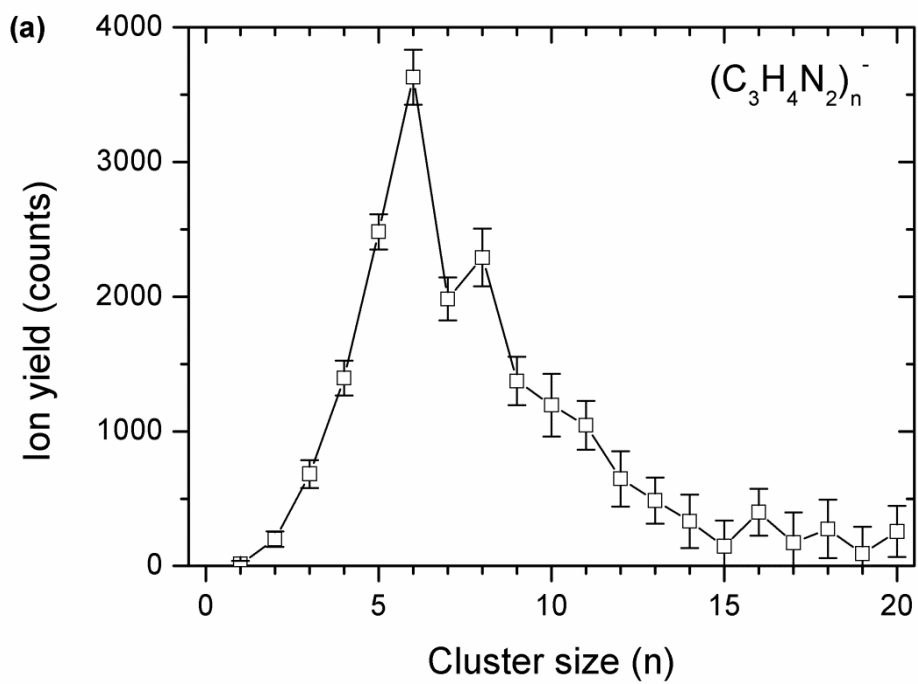

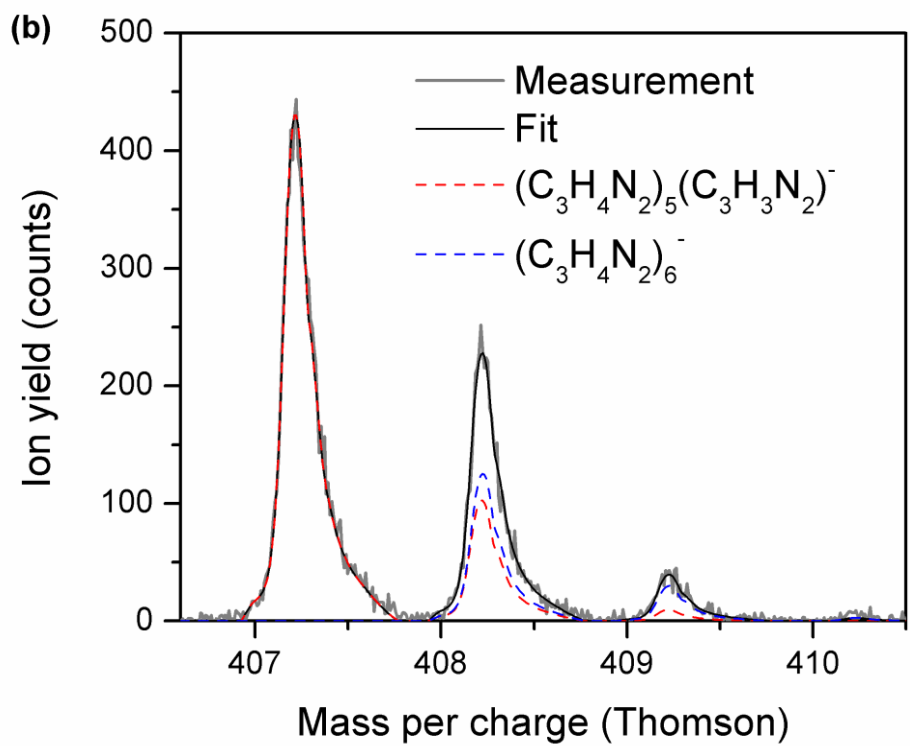

**Figure 5**

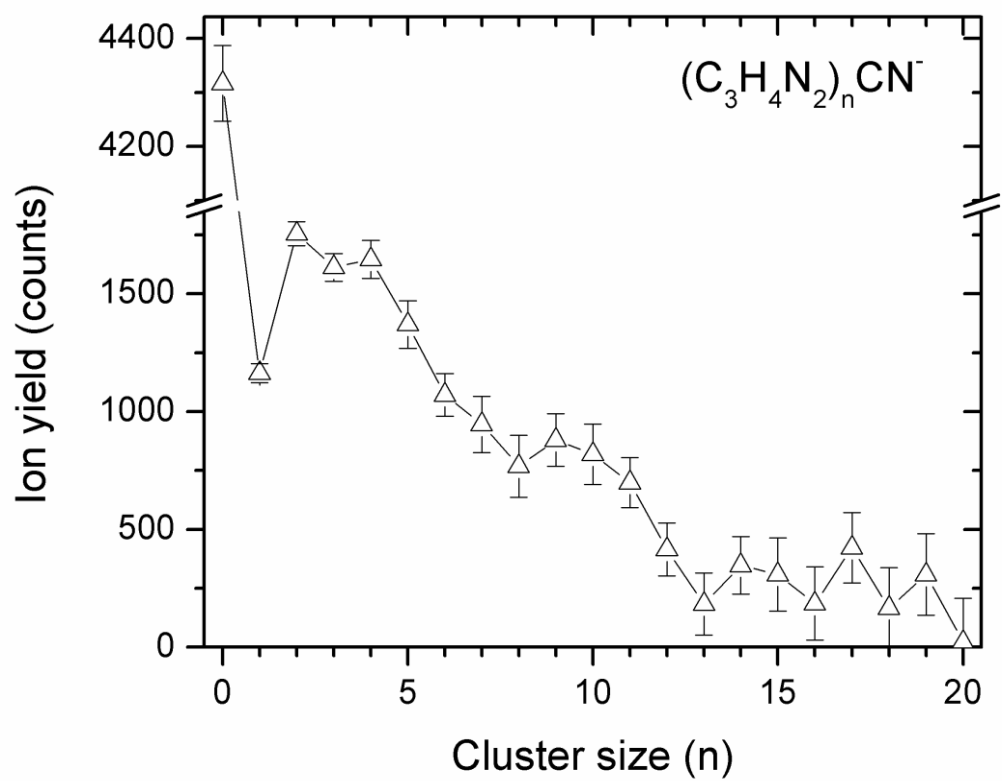

**Figure 6**

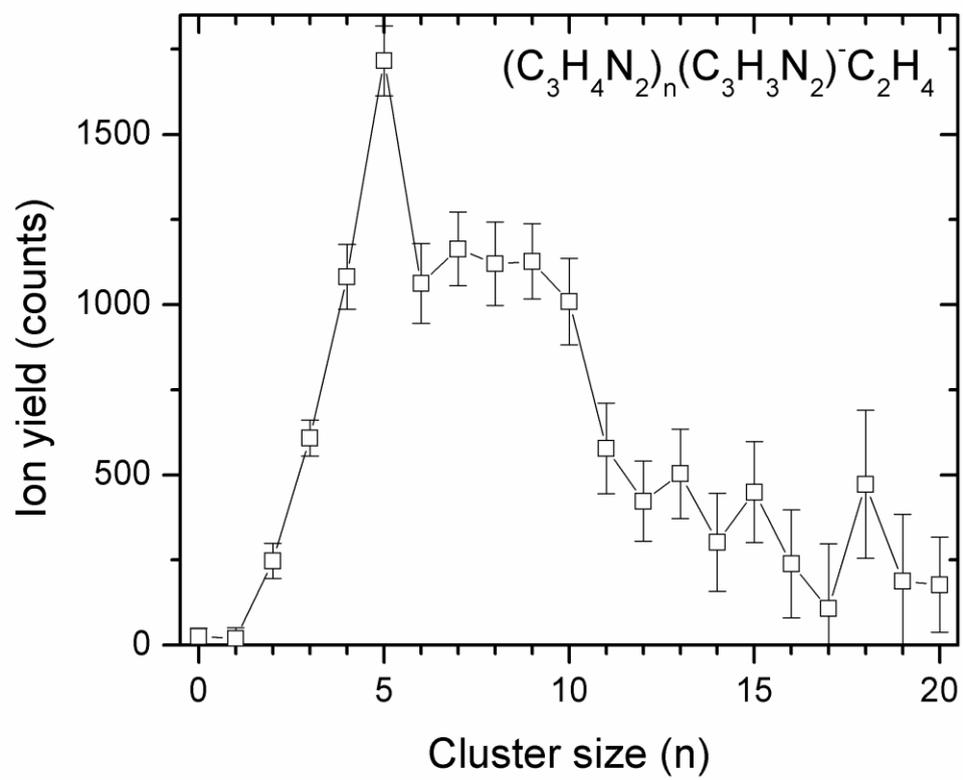

**Figure 7**

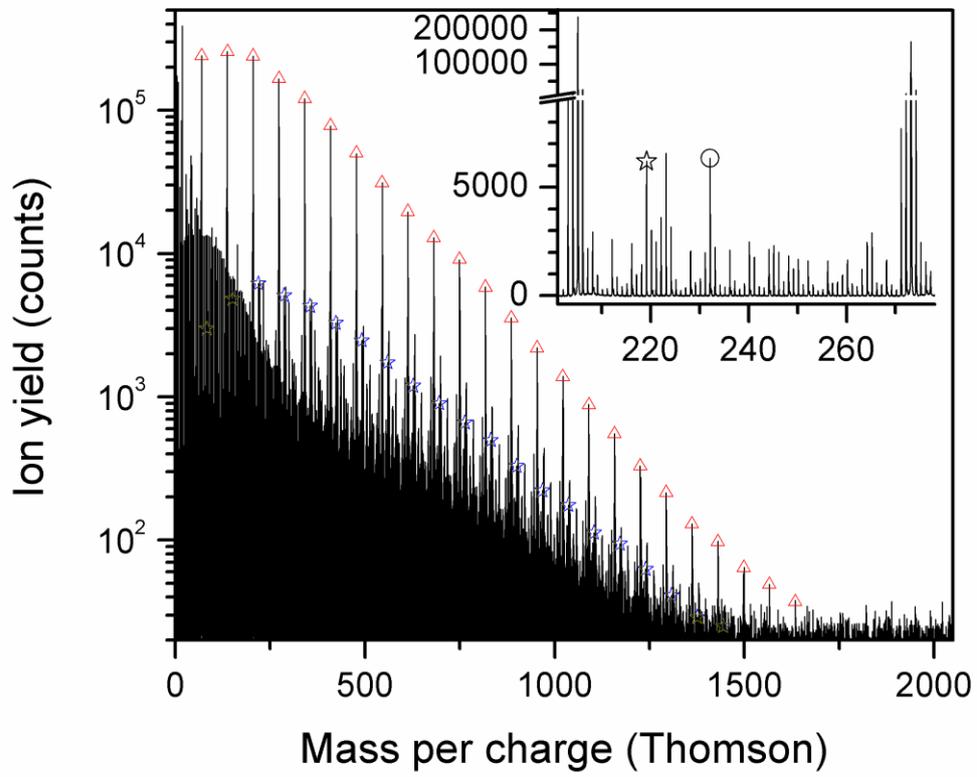

**Figure 8**

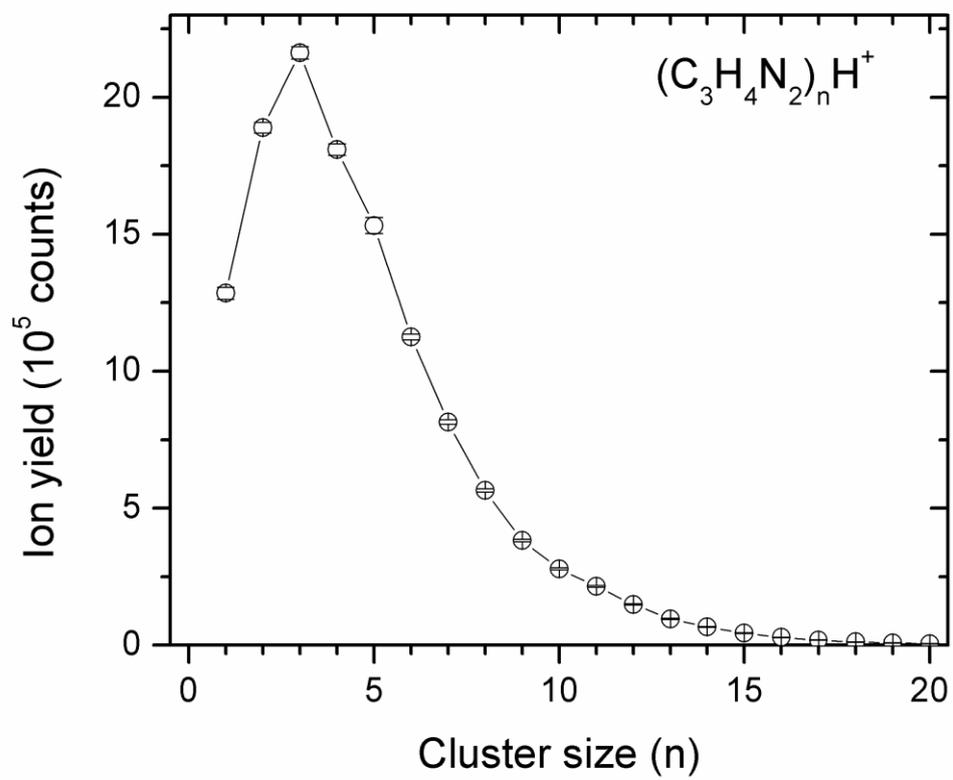

**Figure 9**

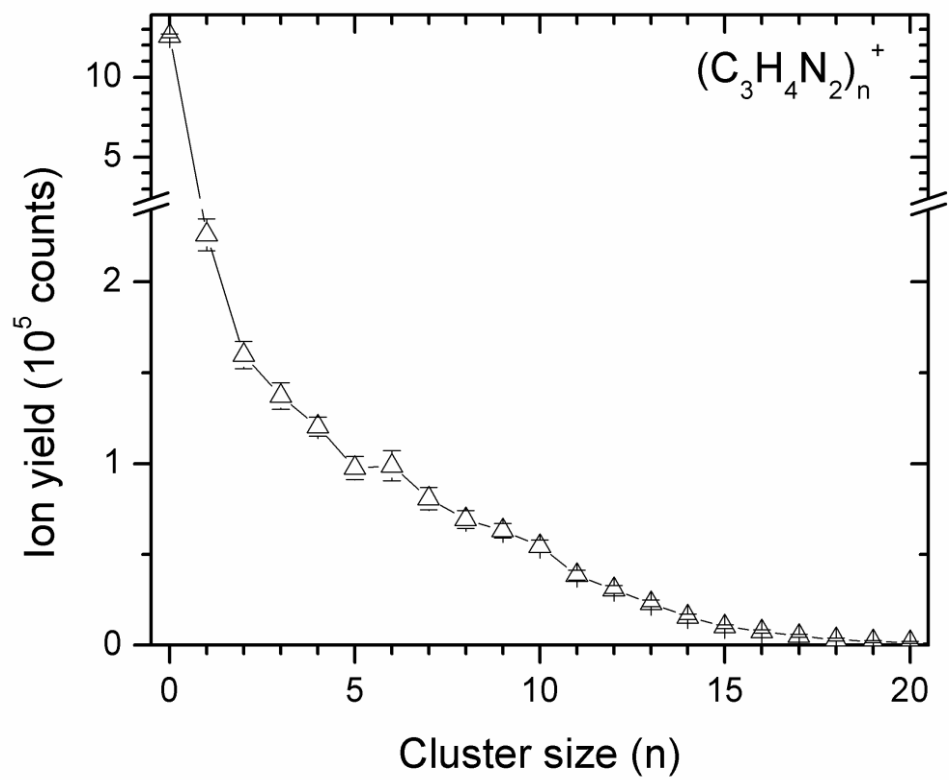

**Figure 10**

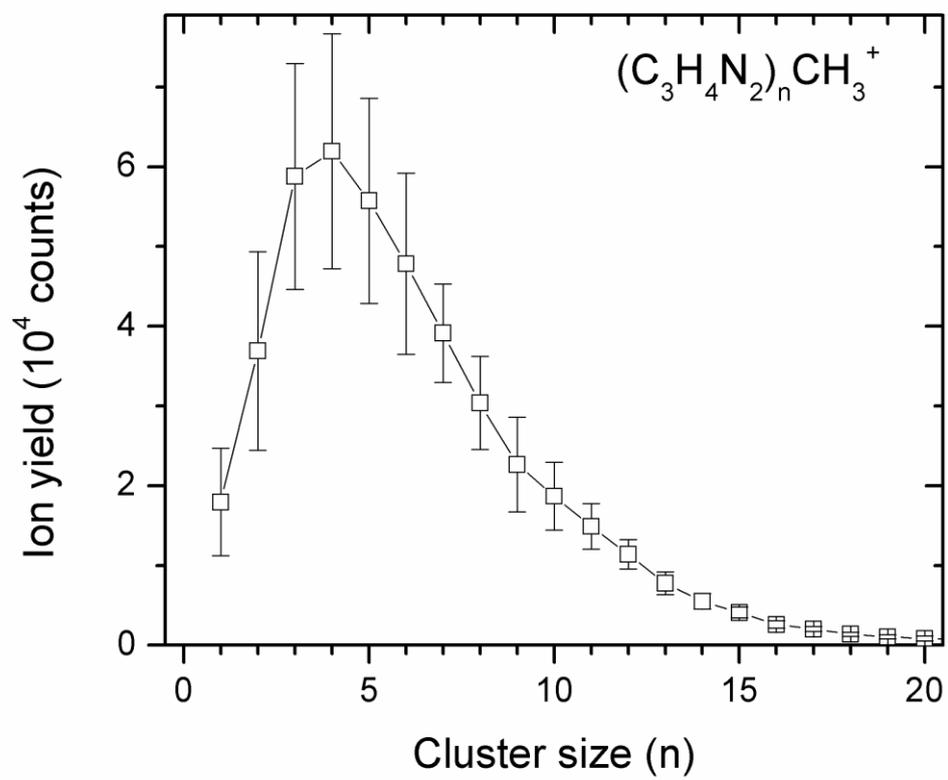